\documentclass[12pt]{elsart}

\usepackage{amssymb}
\usepackage{graphicx}
\usepackage{tabularx}

\providecommand{\Journal}[4]{ {#1} {#2} (#4) #3}
\providecommand{\A}{Auditing} %
\providecommand{\ARAA}{Annu. Rev. Astron. Astr. } %
\providecommand{\AJP}{Am. J. Phys.} %
\providecommand{\AJ}{Astron. J. } %
\providecommand{\AJM}{Am. J. Math.} %
\providecommand{\AMM}{Am. Math. Mon.} %
\providecommand{\AS}{Am. Stat.} %
\providecommand{\C}{Computing} %
\providecommand{\CSSC}{Commun. Stat.-Simul. C.} %
\providecommand{\DCDS}{Discrete Cont. Dyn. S.} %
\providecommand{\EJP}{Eur. J. Phys.} %
\providecommand{\EPJA}{Eur. Phys. J. A} %
\providecommand{\IPCE}{J. Inform. Process. Cybern. EIK} %
\providecommand{\jC}{Chaos} %
\providecommand{\jAS}{Am. Sci.} %
\providecommand{\jSS}{Stat. Sci.} %
\providecommand{\JTP}{J. Theor. Probab.} %
\providecommand{\jAMS}{Ann. Math. Stat.} %
\providecommand{\JATA}{J. Am. Tax. Assoc.} %
\providecommand{\JA}{J. Accountancy} %
\providecommand{\MPLA}{Mod. Phys. Lett. A} %
\providecommand{\PAPS}{Proc. Am. Phil. Soc.} %
\providecommand{\PAMC}{P. Am. Math. Soc.} %
\providecommand{\PRSA}{Proc. R. Soc. A} %
\providecommand{\PPNP}{Prog. Part. Nucl. Phys.} %
\providecommand{\PA}{Physica A} %
\providecommand{\SPA}{Stoch. Proc. Appl.} %
\providecommand{\TAMS}{T. Am. Math. Soc.} %

\journal{Astroparticle Physics, published 33 (2010) 255-262.~~}

\begin{document}

\begin{frontmatter}

\title{Empirical Mantissa Distributions of Pulsars}

\author{Lijing Shao} and
\author{Bo-Qiang Ma}\ead{mabq@phy.pku.edu.cn}

\address{School of Physics and State Key Laboratory of Nuclear
Physics and Technology, \\Peking University, Beijing 100871, China}

\begin{abstract}
The occurrence of digits one through nine as the leftmost nonzero
digit of numbers from real world sources is often not uniformly
distributed, but instead, is distributed according to a logarithmic
law, known as Benford's law. Here, we investigate systematically the
mantissa distributions of some pulsar quantities, and find that for
most quantities their first digits conform to this law. However, the
barycentric period shows significant deviation from the usual
distribution, but satisfies a generalized Benford's law roughly.
Therefore pulsars can serve as an ideal assemblage to study the
first digit distributions of real world data, and the observations
can be used to constrain theoretical models of pulsar behavior.
\end{abstract}


\end{frontmatter}

\clearpage

\section{Introduction}

Pulsars, celestial lighthouses in the sky, are excellent natural
laboratories for the research of fundamental properties of matter
under the circumstances of strong gravity, strong magnetic field,
high density, and extremely relativistic condition. Pulsar physics
has been a forefront field of both astronomy and physics for more
than 40 years since its discovery. Though exciting progresses in
this domain have significantly enlarged our knowledge of
astronomical environment and physical processes, there still remain
outstanding problems~\cite{gs06araa,wnrs07ppnp}. Nowadays, due to
the efforts made by modern observational instruments covering over
various wavelength scopes, \emph{e.g.}, ROSAT, BeppoSAX, Chandra,
XMM-Newton, HST, and VLT, people accumulated a sea of pulsar data,
and the data are accreting significantly all these days. Hence data
analysis and statistical synthesis become a vital step to
characterize pulsar properties and reveal their inner
regularities~\cite{l04,bkps07}.

In this paper, we perform a systematic investigation of the mantissa
distributions of pulsars for the first time. The mantissa $m \in
(-1,-0.1~]\cup [~0.1,~1)$ is the significant part of a
floating-point number $x$, defined as $x = m \times 10^n$, where $n$
is an integer. In this paper, if not noted explicitly, we always
postulate that the numbers are positive for succinct statement.

One might presume that the mantissas of any randomly chosen data set
are approximately uniformly distributed, but that is not the case in
real world. Instead, as stated by Newcomb~\cite{n81ajm}, \emph{``the
law of probability of the occurrence of numbers is such that all
mantissae of their logarithms are equally likely''}.
Subsequently, it leads to the conclusion that the first significant
digit, i.e., $1,2,...,9$, of mantissa is logarithmically
distributed, where the number $1$ appears almost seven times more
often than that of the number $9$. The probability of the occurrence
of the first digit can be expressed in an analytical formula, called
Benford's law~\cite{b38paps} after the name of its second
discoverer,
\begin{equation}\label{benford}
P(k) = \log_{10} (1 + \frac{1}{k}), \, k=1,2,...,9
\end{equation}
where $P(k)$ is the probability of a number having the first digit $k$.

Empirically, the areas of lakes, the lengths of rivers, arabic
numbers on the front page of a newspaper~\cite{b38paps}, physical
constants and distributions~\cite{bk91ajp,lm}, the stock market
indices~\cite{l96as}, file sizes in a personal
computer~\cite{tfgs07ejp}, survival distributions~\cite{lse00as},
widths of hadrons~\cite{sm09mpla}, even dynamical
systems~\cite{tbl00c,bbh05tams,b05dcds}, conform to the peculiar law
well. Nevertheless, there also exist other types of data, e.g.,
lottery and telephone numbers, which do not obey the law.
Unfortunately, there is no a priori criteria yet to judge which type
a data set belongs to. In practice, the law is already applicable in
distinguishing and ascertaining fraud in taxing and
accounting~\cite{n96jata,nm99a,rr03ja,gw04cssc}, and speeding up
calculation and minimizing expected storage space in computer
science~\cite{bb85c,s88ipce,bhs07amm}.

Since its second discovery in 1938, many attempts have been tried to
explain the underlying reason for Benford's law. For theoretical
reviews, see papers written by Raimi~\cite{r69amm,r76amm,r85paps}
and Hill~\cite{h95amm,h95pamc,h95ss,h98as,hs05spa}. Nowadays, many
breakthrough points have been achieved in this domain, though, there
still lacks a universally accepted final answer. In mathematics,
Benford's law is the only digit law that is
scale-invariant~\cite{bhm08jtp}, which means that the law does not
depend on any particular choice of units, discovered by
Pinkham~\cite{p61ams}. Also Benford's law is
base-invariant~\cite{h95amm,h95pamc,h95ss}, which means that it is
independent of the base $d$ you use. In the binary system ($d$=2),
octal system ($d$=8), or other base system, the data, as well as in
the decimal system ($d$=10), all fit the general first digit law,
$P(k) = \log_{d}(1+1/k)$, $k \in \{1, 2, ..., d-1\}$. Theoretically,
Hill proved that \emph{``scale-invariance implies
base-invariance''}~\cite{h95amm} and \emph{``base-invariance implies
Benford's law''}~\cite{h95pamc} mathematically in the framework of
probability theory. He also proved that random entries, picked from
random distributions, form a sequence whose significant digit
distribution approaches to Benford's law~\cite{h95ss}.

Intending to uncover some regularities of pulsar data, and also to
explore new domains of the digit law, we investigate the mantissa
distributions of pulsar quantities systematically. We find that the
first nonzero digit
of mantissas displays unevenness according to the logarithmic law.
The exceptions are the barycentric period and rotation frequency,
which show significant deviations from Benford's law.
Further, we also discuss various properties of the digit law, and
perform the generalized Benford's law to data sets of barycentric
period and rotation frequency of pulsars as well. Therefore the data
of pulsars provide an ideal assemblage for further studies on the
first digit law of the nature.

\section{First digit distributions}

We investigate the famous Australia Telescope National Facility
(ATNF) pulsar
catalogue\footnote{http://www.atnf.csiro.au/research/pulsar/psrcat},
which is a widely used database, maintained by Manchester \emph{et
al.}~\cite{mhth05aj}. Thanks to their exhaustive search of pulsar
literatures, at least back to 1993, data from all papers announcing
the discoveries of pulsars or giving improved parameters are entered
into the catalogue database. As ATNF pulsar catalogue is an updating
database, to avoid ambiguity, in this paper we make use of
\emph{Catalogue version 1.36}, including totally 1826 samples of
pulsars. However, not every pulsar item has complete information,
e.g., one of 1826 pulsars, J1911+00, in this database has no
information on barycentric period hence rotation frequency. We
cover every presented nonzero number without bias. The notations of
physical quantities in the following text are the same as that from
this catalogue.

\begin{figure}
\begin{center}
\includegraphics[scale=1]{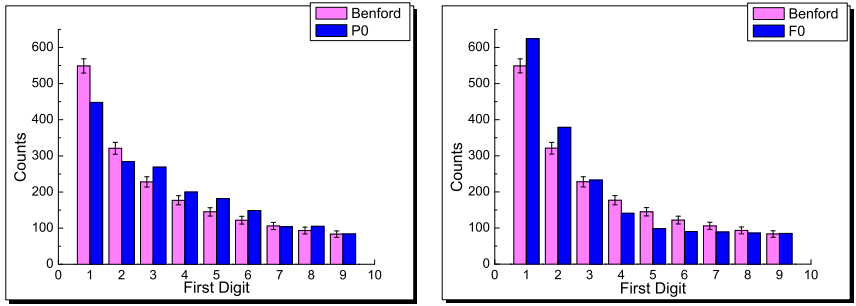}
\caption{Comparisons of Benford's law and the distributions of the
first digit of the barycentric period (left) and rotation frequency (right) of
pulsars.\label{gpf}}
\end{center}
\end{figure}

\begin{table}
\begin{center}
\caption{The first digit distributions of the period, frequency, spin down ages, and
their time derivatives of pulsars.\label{tpf}}
\begin{tabularx}{14cm}{Xcccc}
\hline\hline
Physical Quantity & Notation & Data points & $\chi^2(8)$ & $p$-value\\
\hline
Barycentric period of the pulsar (s)    &   P0  &   1825    &
50.552 & 0.0001\\
Barycentric rotation frequency (Hz) &   F0  &   1825    &   54.577 & 0.0001\\
Time derivative of barycentric period (dimensionless)  &   P1  &
1695    &   4.497 & 0.8097\\
Period derivative corrected for proper motion effect    &   P1\_i &
219 &   9.502 & 0.3017\\
Time derivative of barycentric rotation frequency (${\rm s}^{-2}$) &
F1  &   1695    &   7.539 & 0.4797\\
Second time derivative of barycentric rotation frequency (${\rm s}^{-3}$)  &
F2  &   395 &   3.020 & 0.9331\\
Spin down age (yr) [$\tau = P/(2\dot{P})$]  &   Age &   1664    &
3.721 & 0.8814 \\
Spin down age from P1\_i (yr)    &   Age\_i   &   219 &   7.078 & 0.5282\\
\hline
\end{tabularx}
\end{center}
\end{table}

The first digit distributions of the barycentric period in the unit
of second, P0, and rotation frequency in the unit of Hertz, F0, of
pulsars are illustrated in Figure~\ref{gpf}  and Table~\ref{tpf}.
There are totally $N = 1825$ available samples as mentioned. The
theoretical predictions in the figure are the expected number,
$N_{\rm Ben} = N \, \log_{10}(1+1/k)$,
together with the root mean square error evaluated by the binomial
distribution~\cite{bmp93ejp,nr08epja}, $\Delta N = \sqrt{N \, P(k)
\,(1-P(k))}$.
We use the Pearson $\chi^2$ to estimate goodness of fit to the
probability distribution,
\begin{equation}\label{pearson}
\chi^2(n-1) = \sum_{i=1}^{n} \frac{(N_{\rm Obs} - N_{\rm
Ben})^2}{N_{\rm Ben}}
\end{equation}
where $N_{\rm Obs}$ is the observational number and $N_{\rm Ben}$ is
the theoretical prediction from Benford's law, and here in our
question $n=9$. In Eq.~(\ref{pearson}), the degree of freedom is
$9-1=8$, and under the confidence level (CL) 95\%, $\chi^2(8) =
15.507$.

From Figure~\ref{gpf}, we see that they are roughly consistent but
with considerable divergence from the Benford's law. The deviations
are actually just one deviation, since period and frequency are
inversely related. Pearson $\chi^2$ in Table~\ref{tpf}
are rather large, thus the $\chi^2$ tests
reject the null hypothesis $H_0$: \emph{the barycentric period and
rotation frequency of pulsars fit Benford's law}. Worthy to mention
that, their $p$-values, which measure the probability of obtaining a
test statistic at least as extreme as the one that is actually
observed, are rather small accordingly. The lower the $p$-value, the
less likely the null hypothesis. In our analysis, we reject a null
hypothesis if its $p$-value is less than $0.05$ (equivalent to under
${\rm CL}~95\%$). The anomaly of period and frequency from Benford's
law will be discussed and handled with the generalized Benford's law
latter in the paper.

\begin{figure}
\begin{center}
\includegraphics[scale=1]{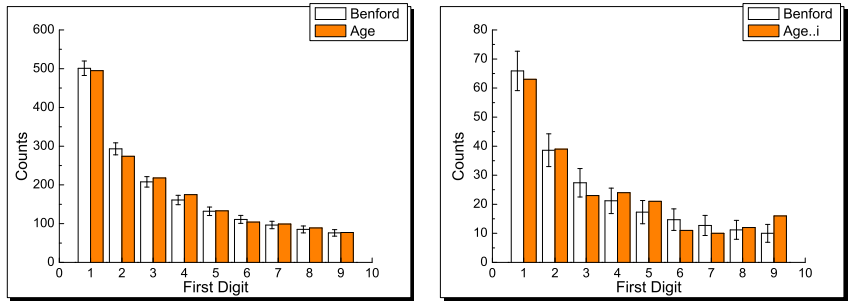}
\caption{Comparisons of Benford's law and the distributions of the
first digit of the spin down age (left) and the spin down age
from P1\_i (right) of
pulsars.\label{gage}}
\end{center}
\end{figure}

However, as depicted in Figure~\ref{gage} and Table~\ref{tpf},
comparisons of Benford's law and the distributions of the first
digit of the spin down ages of pulsars, Age and Age\_i, are both
very impressive. The ages are estimated as $\tau = P/(2\dot{P})$,
where $P=P0$, and $\dot{P} = P1$ for Age while $\dot{P} = P1\_{\rm
i}$ for Age\_i. As listed in Table~\ref{tpf}, P1 and P1\_i are
observational time derivative of barycentric period and the period
derivative corrected for proper motion effect respectively. The
$\chi^2$
strongly supports the null hypothesis $H_0$: \emph{the spin down
ages of pulsars follow Benford's law}. Torres \emph{et
al.}~\cite{tfgs07ejp} discussed the possibility of transformation
from non-Benford data sets to Benford ones. They started from evenly
distributed random numbers, and found that the group of numbers
created by random multiplication between them reveals uneven
property and fulfils the logarithmic law. Correspondingly, we here
give an empirical demonstration from a non-Benford set (the
barycentric periods) and a Benford set (time derivatives of
barycentric period) to a Benford one (the spin down ages).

\begin{figure}
\begin{center}
\includegraphics[scale=1.4]{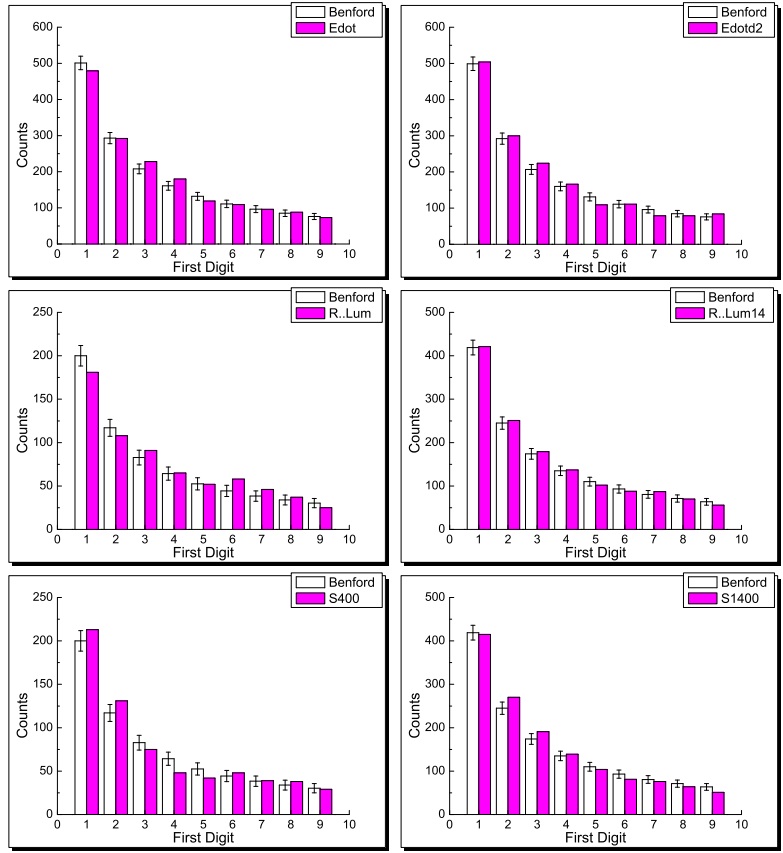}
\caption{Comparisons of Benford's law and the distributions of the first
digit of the power
quantities of pulsars. From top left to bottom right, they are spin down
energy loss rate, energy flux at the Sun, radio luminosity at 400 MHz,
radio luminosity at 1400 MHz, mean flux density at 400 MHz, and mean flux
density at 1400 MHz.\label{gpower}}
\end{center}
\end{figure}

\begin{table}
\begin{center}
\caption{The first digit distributions of the power
quantities of pulsars.\label{tpower}}
\begin{tabularx}{14cm}{Xcccc}
\hline\hline
Physical Quantity & Notation & Data points & $\chi^2(8)$ & $p$-value\\
\hline
Spin down energy loss rate (ergs/s) &   Edot    &   1664    &
6.601 & 0.5802\\
Energy flux at the Sun (ergs/${\rm kpc}^2$/s)   &   Edotd2  &   1656
&   9.938 & 0.2694\\
Radio luminosity at 400 MHz \mbox{(mJy ${\rm kpc}^2$)} &   R\_Lum  &   663
&   10.083 & 0.2592\\
Radio luminosity at 1400 MHz \mbox{(mJy ${\rm kpc}^2$)}    &   R\_Lum14 &
1391    &   2.673 & 0.9532\\
Mean flux density at 400 MHz (mJy)  &   S400    &   663 &   10.446 & 0.2351\\
Mean flux density at 1400 MHz (mJy) &   S1400   &   1391    &
9.855 & 0.2754\\
\hline
\end{tabularx}
\end{center}
\end{table}

The first digit distributions of power quantities of pulsars are
depicted in Figure~\ref{gpower} and Table~\ref{tpower}. In
Figure~\ref{gpower}, from top left to bottom right, they are spin
down energy loss rate, energy flux at the Sun, radio luminosity at
400 MHz, radio luminosity at 1400 MHz, mean flux density at 400 MHz,
and mean flux density at 1400 MHz.
The
$\chi^2$ tests for all the power quantities
are well supportive to $H_0$: \emph{the power quantities
of pulsars fit Benford's law}.

\begin{figure}
\begin{center}
\includegraphics[scale=1.4]{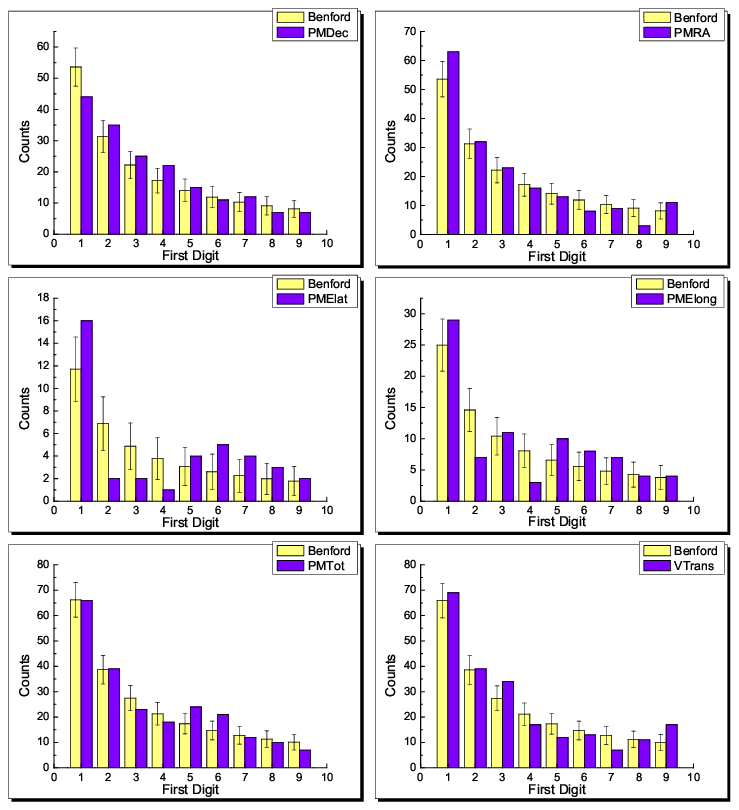}
\caption{Comparisons of Benford's law and the distributions of the first
digit of the kinematic
quantities of pulsars. From top left to bottom right, they are proper motion
in declination, proper motion in right ascension, proper motion in ecliptic
latitude, proper motion in ecliptic longitude, total proper motion, and the
transverse velocity based on the best estimated pulsar distance.\label{gmove}}
\end{center}
\end{figure}

\begin{table}
\begin{center}
\caption{The first digit distributions of the kinematic
quantities of pulsars.\label{tmove}}
\begin{tabularx}{14cm}{Xcccc}
\hline\hline
Physical Quantity & Notation & Data points & $\chi^2(8)$ & $p$-value\\
\hline
Proper motion in declination (mas/yr)   &   PMDec   &   178 &
4.840 & 0.7745\\
Proper motion in right ascension (mas/yr)   &   PMRA    &   178 &
8.421 & 0.3935\\
Proper motion in ecliptic latitude (mas/yr) &   PMElat  &   39  &
13.057 & 0.1099\\
Proper motion in ecliptic longitude (mas/yr)    &   PMElong &   83 &
11.695 & 0.1653\\
Total proper motion (mas/yr)    &   PMTot   &   220 &   7.527 & 0.4810\\
Transverse velocity -- based on DIST (km/s)  &   VTrans  &   219 &
11.855 & 0.1578\\
\hline
\end{tabularx}
\end{center}
\end{table}

Likewise, in Figure~\ref{gmove} and Table~\ref{tmove}, we present
the first digit distributions of kinematic quantities of pulsars. In
Figure~\ref{gmove}, there are proper motion in declination, proper
motion in right ascension, proper motion in ecliptic latitude,
proper motion in ecliptic longitude, total proper motion, and the
transverse velocity based on the best estimated pulsar distance,
from top left to bottom right.
The $\chi^2$ tests for all kinematic quantities, the total proper
motion as well as components of the total motion, favor the null
hypothesis $H_0$: \emph{the movements of pulsars conform to
Benford's law}. 

\begin{figure}
\begin{center}
\includegraphics[scale=1]{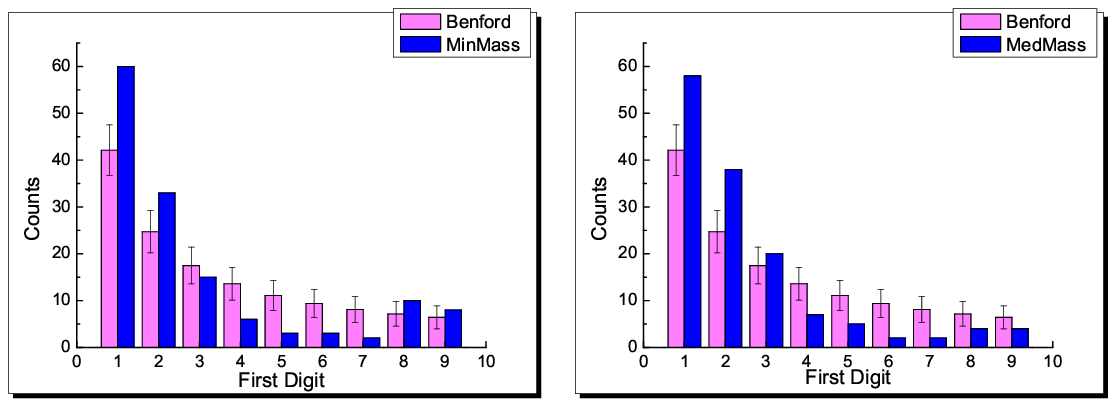}
\caption{Comparisons of Benford's law and the distributions of the
first digit of the expected minimum companion mass (left) and
median companion mass (right) of
pulsars.\label{gmass}}
\end{center}
\end{figure}

\begin{table}
\begin{center}
\caption{The first digit distributions of the expected minimum companion mass
and median companion mass of pulsars.\label{tmass}}
\begin{tabularx}{14cm}{Xcccc}
\hline\hline
Physical Quantity & Notation & Data points & $\chi^2(8)$ & $p$-value\\
\hline
Minimum companion mass assuming i=90 degrees and neutron star
mass
is 1.35 ${\rm M}_{\odot}$  &   MinMass &   140 &   31.331 & 0.0001\\
Median companion mass assuming i=60 degrees &   MedMass &   140 &
32.781 & 0.0001\\
\hline
\end{tabularx}
\end{center}
\end{table}

As mentioned above, not all data sets respect Benford's law.
Especially, artificial numbers often diverge from the logarithmic
distribution, and this is the very reason why Benford's law is
applied in detecting number frauds. Data sets that are arbitrary and
contain restrictions usually do not comply with the peculiar law, in
contrast, data sets measured from the real world are more likely to
obey the law. The first digit distributions of the expected
companion minimum mass, MinMass, and median companion mass, MedMass,
of pulsars are illustrated in Figure~\ref{gmass} and
Table~\ref{tmass}. MinMass is the minimum companion mass calculated
by assuming $i=90$~degrees and neutron star mass is 1.35~${\rm
M}_{\odot}$, and MedMass is the median companion mass assuming
$i=60$ degrees, where $i$ denotes the orbital obliquity. Apart from
the feasibility of the rough assumptions, the expected minimum mass
and median mass are neither natural numbers by measurement, and the
physics of these companion masses also places constraints on the
range of values, making it difficult for their distributions to be
logarithmically uniform, thus the departures from the logarithmic
distribution with rather large $\chi^2$
for MinMass and
MedMass are expected. However, in general, it still lacks universal
justification for all examples.

\begin{figure}
\begin{center}
\includegraphics[scale=1]{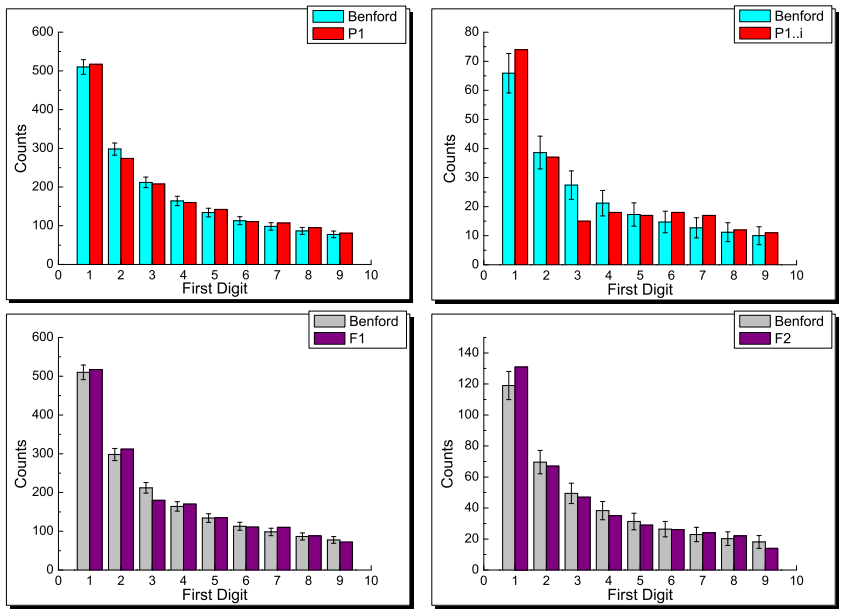}
\caption{Comparisons of Benford's law and the distributions of the
first digit of the time derivatives of period and frequency of
pulsars. From top left to bottom right, they are time derivative of
barycentric period, period derivative corrected
for proper motion effect, time derivative of barycentric rotation
frequency, and second time
derivative of barycentric rotation frequency.\label{gpfs}}
\end{center}
\end{figure}

As we can see at the beginning of this section, the barycentric
period and rotation frequency are not good to follow Benford's law,
hence it raises an interesting question: do the time derivatives of
these non-Benford data sets follow the significant digit law? In
Figure~\ref{gpfs}, we present an empirical positive answer to the
above question. From Table~\ref{tpf}, we can see that the Pearson
$\chi^2$ for time derivative of barycentric period, period
derivative corrected for proper motion effect, time derivative of
barycentric rotation frequency, and second time derivative of
barycentric rotation frequency, 
drive the null hypothesis $H_0$: \emph{the time derivatives of
period and frequency of pulsars obey Benford's law}, to be accepted.

Furthermore, we investigate the positive and negative numbers
separately for this case, and this has not been discussed explicitly
in other literatures. We find that the Pearson $\chi^2$ for the 1664
positive numbers and 31 negative ones of the time derivative of
barycentric period of pulsars are 5.364 and 9.977, while for the
time derivative of barycentric rotation frequency, 31 positive and
1664 negative numbers, $\chi^2$ are 2.813 and 7.590, respectively.
Thus we hint at the conclusion that the negative numbers, as well as
positive ones, empirically apply to Benford's law. It is a challenge
why the nature spontaneously organizes numbers into such a fantastic
global order.

\section{Generalized Benford's law}

\begin{figure}
\begin{center}
\includegraphics[scale=1.1]{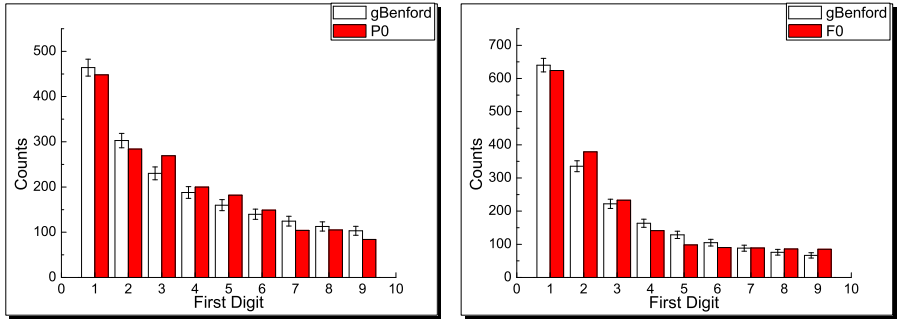}
\caption{Comparisons of the generalized Benford's
law and the distributions of the
first digit of the barycentric period (left) and
rotation frequency (right) of
pulsars.\label{general}}
\end{center}
\end{figure}

Pietronero \emph{et al.}~\cite{pttv01pa} provided a
new insight, suggesting that a process or an object $m(t)$ with its
time evolution governed by multiplicative fluctuations generates
Benford's law naturally, and they used stockmarket as a convictive
example. The main idea is that $m(t + \delta t) =
r(t) \times m(t)$, where $r(t)$ is a random variable. After treating
$\log r(t)$ as a new random variable, it is a Brownian process $\log
m(t + \delta t) = \log r(t) + \log m(t)$ in the logarithmic space.
Utilizing the central limit theorem in a large sample, $\log m(t)$
becomes uniformly distributed. Thus,
\begin{equation}\label{multi}
P(k) = \frac{\int_{k}^{k+1} {\rm d} \log m(t)}{\int_{1}^{10} {\rm d}
\log m(t)} = \frac{\int_{k}^{k+1} m^{-1} {\rm d} m }{\int_{1}^{10}
m^{-1} {\rm d} m} = \log_{10} (1 + \frac{1}{k}),
\end{equation}
which is exactly the formula of Benford's law given in
Eq.~(\ref{benford}). Later, the idea is well extended to affine
processes $m(t + \delta t) = r(t) \times m(t) + r^\prime(t)$ by
Gottwald and Nicol~\cite{gn02pa} using techniques from ergodic
theory.

From Eq.~(\ref{multi}) above, we can see that the probability
density of a number owning mantissa $m$ is proportional to $m^{-1}$,
a fact which was also gained by Lemons~\cite{l86ajp} in terms of a
probabilistic model of partitioning a conserved quantity. Further,
Pietronero \emph{et al.}~\cite{pttv01pa} generalized the probability
density to be proportional to $m^{-\alpha}$, conserving the
scale-invariant property, thus the probability to have the first
digit $k$ becomes $P_{\alpha}(k) = C \int_{k}^{k+1} m^{-\alpha} {\rm
d} m$, where $C$ is a normalization factor. In the framework of
generalized Benford's law, when $\alpha = 0$, it becomes the uniform
distribution, while $\alpha = 1$ corresponds to Benford's law, and
for increasing $\alpha$, the first digit $k = 1$ appears more
frequent, raising the unevenness of the digit distribution.
Moreover, they analyzed the southern California catalogue for
earthquake magnitudes with $\alpha \simeq 2$ and got rather good
fitness. Recently, Luque and Lacase~\cite{ll09prsa} utilized
size-dependent exponent $\alpha (N)$ to analyse consequences of
primes and Riemann zeta zeros.

Here we treat the non-Benford data sets of the barycentric period
and rotation frequency of pulsars with the generalized Benford's
law.
The results are illustrated in Figure~\ref{general}, where we adopt
$\alpha = 0.8$ for barycentric period and $\alpha = 1.2$ for
rotation frequency. Here the symmetry of exponents about 1 is
expected, since two quantities are related reciprocally. Through
comparisons between Figure~\ref{gpf} and Figure~\ref{general}, we
see clearly that the goodness of fit improves a lot. For the
generalized form, the fitness $\chi^2$ is deduced to $\chi^2(7) =
20.187$ for barycentric period and $\chi^2(7) = 25.569$ for rotation
frequency. $F$-test gives $p=0.0142$ and $p=0.0258$ respectively,
which rejects the null hypothesis that {\it no significant
improvement is introduced after introducing $\alpha$}. Hence the
generalized Benford's law is favored for barycentric period and
rotation frequency of pulsars, the $\chi^2$ is still a bit large
though.

Why and how do overwhelming majority of numbers develop into inverse
probability density hence Benford's law? What is the inner
regularity that distinguishes some generalized power-law anomaly
from the specific logarithmic law? There still remain crucial
challenges. Nevertheless, we can always benefit from regularities
provided by the nature, and use Benford's law to inspect scientific
and social data falsify, and also to test the validity of
theoretical models. We hope to introduce pulsars data as a good
assemblage for such kind of studies.

\section{Summary}

In this paper, we present systematic analysis on the first digit
distributions of mantissas of most fundamental quantities of
pulsars, including barycentric period and rotation frequency,
together with their time derivatives, and many more. The results
reveal obvious departures from the uniform distribution, and small
digits are more prevalent than large ones according to a logarithmic
formula, called Benford's law. However, not all data sets conform to
it. Artificial and restricted data sets often diverge from the law,
such as the expected minimum and median companion masses of pulsars
in our study. Furthermore, 
the generalized Benford's law is applied to the barycentric period
and rotation frequency of pulsars.

Since the first discovery of the peculiar significant digit law more
than one hundred years ago, many crucial understandings of the inner
regularities from mathematical viewpoint have been achieved. But
there still remain big challenges to scientists to look into more
profound physical reasons of such remarkable property of the nature.
Thus the revelation of the hitherto unnoticed regularities in the
pulsar data opens a new window to look into novel aspects of
astronomical objects.

\section*{Acknowledgments}

This work is partially supported by National Natural Science
Foundation of China (Nos.~10721063, 10975003) and National Fund for
Fostering Talents of Basic Science (Nos.~J0630311, J0730316). It is
also supported by Hui-Chun Chin and Tsung-Dao Lee Chinese
Undergraduate Research Endowment (Chun-Tsung Endowment) at Peking
University.

\end{document}